\documentclass[preprint,aps, preprintnumbers, nofootinbib]{revtex4}
\usepackage[dvips]{graphics}

\newcommand{\lsim}[1]{
\setlength{\unitlength}{12pt}
\begin{picture}(1.4,1.)
\put(.7,-0.3){\makebox(0.0,1.)[t]{$<$}}
\put(.7,-0.3){\makebox(0.0,1.)[b]{$\sim$}}
\end{picture}#1}

\newcommand{\gsim}[2]{
\setlength{\unitlength}{12pt}
\begin{picture}(1.4,1.)
\put(.7,-0.3){\makebox(0.0,1.)[t]{$>$}}
\put(.7,-0.3){\makebox(0.0,1.)[b]{$\sim$}}
\end{picture}#2}

\begin{document}

\title{Holographic bounds and Higgs inflation}

\author{R. Horvat}
\email{horvat@lei3.irb.hr}
\address{Rudjer Bo\v{s}kovi\'{c} Institute, P.O.B. 180, 10002 Zagreb,
Croatia}

\begin{abstract}
In a recently proposed scenario for primordial inflation, where the 
Standard Model (SM) Higgs boson plays a role of the inflation field, an
effective field theory (EFT) approach is the most convenient for working out
the consequences of breaking of perturbative unitarity, caused by the strong
coupling of the Higgs field to the Ricci scalar. The
domain of validity of the EFT approach
is given by the ultraviolet (UV) 
cutoff, which, roughly speaking, should always exceed the Hubble parameter
in the course of inflation. On the other hand, applying the trusted principles
of quantum gravity to a local EFT demands that it should only be used to
describe states in a region larger than their corresponding Schwarschild 
radius, manifesting thus a sort of  UV/IR correspondence. We consider
both
constraints on EFT, to ascertain which models of the SM Higgs inflation are able
to simultaneously comply with them. We also show that 
if the gravitational coupling evolves with 
the scale
factor, the holographic constraint can be alleviated significantly 
with minimal set of canonical assumptions, by forcing
the said coupling to be asymptotically free.
\end{abstract}

\newpage

\maketitle

In search for the most elegant model for primordial inflation, 
keeping at the same time  a minimal
particle content, a recently
proposed model in which the role of the 
inflaton field is played by the Higgs boson of the
SM, certainly leaps out
\cite{1}. The original model, with no further degrees of freedom beyond the
SM ones, requires an unnaturally large coupling between
the Higgs field and the scalar curvature, typically $\xi \sim 10^4$, to
produce successful inflation. However, the same coupling becomes responsible for
the presence of higher dimensional operators (inducing in so doing 
also some scale
$\Lambda$), whose influence could no longer be
ignored at energies near $\Lambda$. Thus, at energies 
$E \simeq \Lambda$  a perturbative
treatment of the inflatory dynamics 
ceases to be valid because of violation  of the tree-level 
unitarity (or, alternatively, because of entering the strong coupling regime). The 
cutoff $\Lambda$
turns out to be dependent of the background value of the Higgs field
\cite{2}, and since it could be quite large during inflation, there is a hope
that the intrinsic energy scale involved in the physical processes during
inflation be less
than $\Lambda$, thereby necessitating no understanding on the
UV completion of the theory. Heuristically, the EFT approach, being thus
insensitive to new physics above $\Lambda$ at tree level \cite{2}, 
suffices if $\Lambda >> H_{inf}$, where $H_{inf}$ is is Hubble parameter during
inflation. The question of whether the above scenario is free from unitarity
problems  has been discussed extensively recently \cite{3, 4, 5, 6, 7, 8, 9,
10, 11}. For earlier attempts, see \cite{12}.

Especially promising in this respect has appeared the new model of the SM Higgs 
inflation, featuring the unique nonminimally derivative coupling of the Higgs
boson to gravity \cite{13}. However, together with the original model
\cite{1}, this
one too was shown to suffer from unitarity problems, meaning
that some new physics is likely to show up in the inflatory dynamics
\cite{14, 15}. Finally, using the Palatini formulation of gravity instead of 
the metric one, the original model of Higgs inflation can be made a UV-safe, 
as the UV cutoff is always larger than the Hubble rate \cite{16}.

In the present paper, we would like to draw attention to another constraint
which could possibly endanger the EFT approach of the SM Higgs inflation -
the holographic bound. Our main observation here is that an EFT underlying
inflation should be valid in the whole Hubble volume during inflation. However, 
whenever EFT is 
to be valid in volumes $>> \Lambda^{-3}$, this might jeopardize the absolute
bound on the entropy of a black hole. Indeed, for an EFT in a box of size
$L$ (providing an IR cutoff) the entropy scales extensively, $S_{EFT} \sim
L^3 \Lambda^3$, and therefore there is always a sufficiently large volume for
which $S_{EFT}$ would exceed the absolute Bekenstein-Hawking bound $S_{BH}
\sim L^2 M_{Pl}^2$, where $M_{Pl}$ is the Planck mass. To prevent such an
undesirable scenario, a kind of UV/IR 
correspondence, $L\Lambda^3 \lsim M_{Pl}^2$ \cite{17},
is to be imposed on a theory to assure its validity in arbitrarily large
volumes. However, at saturation, this bound means that an EFT
should also be capable to describe systems containing black holes, since it
necessarily includes many states with Schwarzschild radius much larger than
the box size. There are however arguments for why an ordinary local EFT
appears unlikely to provide an adequate description of any systems
containing black holes, which we list below.

Having become a trusted principle of quantum gravity and even 
part of the mainstream after the Maldacena's discovery of
AdS/CFT duality \cite{18}, the holographic principle \cite{19, 20} virtually 
deprives 
any local EFT  from being capable to  describe states in a region smaller
than their corresponding Schwarschild radius \cite{17}. For instance, as a    
closely related concept  emerging from the holographic principle, the black hole
complementarity \cite{21} relativizes the very notion of location a certain
phenomenon takes place at, something that an ordinary local EFT certainly is
not capable to describe. On relying on locality, as an important property of
quantum field theory, we get that quantum evolution seemingly ceases to
be unitary in black hole backgrounds (operators at space-like points 
no longer commute) \cite{21}. Also, near the horizon of a
black hole, the momenta of infalling particles as seen by an outside
observer grow exponentially with time \cite{22}, a phenomenon certainly at odds 
with effective, low-energy description. Thus, summing up, ordinary EFTs
may not be valid to describe those states already collapsed to a
black hole, implying a constraint much more stringent  
than that obtained from the entropy of a black 
hole, $\Lambda^4 \lsim L^{-2} M_{Pl}^2$ \cite{17}. \footnote{This relates
to a huge gap in entropy between a system on the brink of experiencing a
sudden collapse to a black hole $(S  \sim S_{BH}^{3/4})$ 
and the black hole itself, with both systems
having the same size and energy \cite{17, 23, 24}.} Note, though, 
that this latter constraint
is not absolute in character as the former one, obtained from the
requirement that for an EFT in an arbitrarily-sized box the entropy should
not exceed the entropy  of the same-sized black hole. Still, it is based on
the generally accepted view 
(supported, among others, with the bunch of strong arguments as given above) 
that black holes and their interaction with
quantum fields are not likely to be described by any EFT. So, we deem this
prevailing wisdom as a solid foundation for the latter, more stringent
constraint, which we are going employ in this paper.    

The EFT approach of inflatory dynamics thus requires that the Hubble length
always exceeds the Schwarschild radius of states  encompassed by the Hubble
volume, $H_{inf}^{-1} \gsim \; R_{S}$, or equivalently \footnote{Throughout this 
paper 
we aim not to keep records of numerical factors (of order O(1)).}
\begin{equation}
\Lambda^4 \lsim H_{inf}^2 \bar{M}_{Pl}^2,
\end{equation}
where $\bar{M}_{Pl}$ is the reduced Planck mass. Note that (1) bounds
$\Lambda$ from above, whereas the (heuristic) constraint coming from
violation of unitarity gives the lower bound for $\Lambda$. In combination,
both constraints require $\Lambda << \bar{M}_{Pl}$. However, we study
below more precisely some Higgs inflation models to see if they comply to (1).
\footnote{The Hubble parameter in (1) gives rise to a
cosmological event horizon and therefore to a cosmological black hole,
which, from our perspective, lies beyond the horizon, leaving thus our
considerations unaffected.}      

The original (toy) model of Higgs inflation \cite{1} is given by the action
\begin{equation}
S = \int d^4 x \sqrt{-g} \left [ \frac{\bar{M}_{Pl}^2}{2}R + \xi
\cal{H}^{\dagger}\cal{H}R + \cal{L}_{SM}^{\rm {s}} \right ]\;,
\end{equation}
where $R$ is the Ricci scalar, $\cal{H}$ is the SM Higgs doublet and
$\cal{L}_{SM}^{\rm {s}}$ is the scalar part of the SM Lagrangian. As already stated, the Higgs-Ricci
coupling of order of $\xi \sim 10^{4}$ is needed to obtain successful
inflation. The background Higgs field during the inflationary 
period, $\bar{\phi} >> \bar{M}_{Pl}\xi^{-1/2}$, gives
rise to an UV cutoff in the Jordan frame $\Lambda^J \simeq
\xi^{1/2}\bar{\phi}$ \cite{2}. On the other hand, the Hubble parameter in
the Jordan frame reads $H_{inf}^J \simeq \lambda^{1/2} \xi^{-1/2}\bar{\phi}$,
where $\lambda$ is the Higgs quartic coupling entering $\cal{L}_{SM}^{\rm
{s}}$.
Plugging all this back into (1), one arrives at $\xi << \lambda^{1/2}$.
Considering the current experimental bounds on $\lambda$ of order of
$\rm{O}(10^{-1})$ \cite{25}, one plainly sees that the original (toy)
model of the SM Higgs inflation does not respect the holographic bound.   

Next, we examine within the same model of the Higgs-driven inflation 
whether the reheating phase
immediately after inflation does respect the holographic bound (1). When
the inflatory epoch relinquishes its dominance to a matter dominated stage
featuring scalar field oscillations, we have $\xi^{-1} \bar{M}_{Pl} <<
\bar{\chi} < \bar{M}_{Pl}$ \cite{2, 26}, where $\chi$ is the 
canonically normalized scalar field in
the Einstein frame. During this stage the scalar potential is
essentially quadratic, and the Hubble parameter is approximately given as
$H_{reh}^E \simeq \lambda^{1/2} \xi^{-1} \bar{\chi}$. With $\Lambda^E \simeq
\bar{\chi}$ \cite{2, 26}, one finds by using (1) 
that $\bar{\chi} \lsim \lambda^{1/2}
\xi^{-1} \bar{M}_{Pl}$. Again, the holographic bound is not respected.

The addition of the gauge bosons in the original (toy) model of Higgs
inflation results in lower cutoffs both in the inflatory and reheating epoch
\cite{2}. In the Einstein frame, the above cutoffs are replaced with
$\Lambda^E \simeq \xi^{-1/2} \bar{M}_{Pl}$ for the inflatory 
epoch, and with
$\Lambda^E \simeq \xi^{-1/2} \bar{\chi}^{1/2} \bar{M}_{Pl}^{1/2}$ 
for the reheating
phase. For the inflatory phase, Eq. (1) gives $\lambda \gsim \; 1$, which means
that even upon inclusion of the SM fields, the holographic bound is not
respected. In the reheating phase $\xi^{-1} \bar{M}_{Pl}
<< \bar{\chi} < \bar{M}_{Pl}$, the requirement (1)  
gives again $\lambda \gsim \; 1$. This shows that the inclusion of the 
gauge bosons does not
reduce the cutoffs sufficiently, so that (1) is not obeyed 
by the original model.  

The unitarity violation scale is defined by the minimal cutoff read out of
all higher dimensional operators. In the original model of Higgs inflation,
the unitarity violation scale in the Jordan frame is given by the cutoff read
out of the lowest (5-dimensional) operator, once one expands about the
background values. It is just the opposite for the
holographic bound, where the cutoff appears as a lower bound in Eq. (1).
This means that inclusion of higher dimensional operators could aggravate the
holographic bound. Indeed, it can be shown that the cutoffs associated to higher
dimensional operators $(n > 5)$ get further amplified  by the background value of the
scalar field, the highest energy scale during inflation. The original model
of Higgs inflation is undoubtedly at odds with the holographic bound. 

The new model of Higgs inflation, promoting the coupling of the
Higgs boson to the Einstein tensor instead to the Ricci scalar (at the same
time keeping the minimal SM content), is given by the action
\cite{13}
\begin{equation}
S = \int d^4 x \sqrt{-g} \left [ \frac{\bar{M}_{Pl}^2}{2}R +
\frac{1}{2}(g^{\mu\nu} - \omega^2
G^{\mu\nu})\partial_{\mu}\phi\partial_{\nu}\phi -\frac{\lambda}{4}\phi^4
\right ]\;,
\end{equation}
where $G^{\mu\nu} = R^{\mu\nu} -(R/2)g^{\mu\nu}$ is the Einstein tensor,
$\phi$ is the real Higgs field and $\omega^{-1}$ is some mass scale. The UV
cutoff of the EFT approach, emerged from violation of the tree-level
unitarity in the course of inflation, is given in the Jordan 
frame as $\Lambda^J \simeq (\bar{H}_{inf}^{J2} \bar{M}_{Pl})^{1/3}$ 
\cite{13, 14}. With $\bar{H} \simeq \lambda^{1/2} \bar{M}_{Pl}^{-1}
\bar{\phi}^2$ \cite{13}, the holographic bound (1) gives $\bar{\phi} \lsim
\lambda^{-4}\bar{M}_{Pl}$. The model has been tested against the most recent 
 WMAP data, to give the allowed size for the background field in a small 
range near $\bar{\phi} \sim 10^{-2} \bar{M}_{Pl}$ \cite{27}. Although it
safely passes the holographic test with the lowest cutoff coming from the
7-dimensional operator, the inclusion of higher dimensional operators could
again vitiate this agreement. Indeed, although with the cutoff read out of
the 8-dimensional operator, $\Lambda^J \simeq (\bar{H}_{inf}^{J2}
\bar{M}_{Pl}^2)^{1/4}$, the holographic bound stays at least marginally
respected, it becomes violated by the posterior higher dimensional operator. The
cutoff read out of 9-dimensional operator, $\Lambda^J \simeq
(\bar{H}_{inf}^{J2} \bar{M}_{Pl}^3)^{1/5}$, implies, when plugged in (1),
$\bar{\phi} \gsim \; \lambda^{-1/4} \bar{M}_{Pl}$. Clearly, this is at odds
with \cite{27}. On top of that, it was
argued recently \cite{14} that the model is still plagued with 
unitarity problems.

Recently, it was shown\cite{16} by using the Palatini instead of 
metric formulation of gravity, that unitarity issues can be alleviated in the
Higgs inflation scenario \cite{1}, retaining at
the same time the minimal particle content. In particular, it was
found that the ratio $\Lambda /H$  
can be made greater in the Palatini formulation of
gravity. However, this would obviously exacerbate compliance with the
holographic bound. 

Another interesting conjecture \cite{14}, 
serving to  increase the ratio $\Lambda /H$,
is to promote the gravitational coupling to a running quantity, with the 
property of being asymptotically safe. Heuristically, while 
the original model \cite{1} is not affected to a great extent since the ratio 
$\Lambda /H$ does not
contain the $\bar{M}_{Pl}$ factor in the inflatory period, the same
ratio in the new
model \cite{13} grows as $\bar{M}_{Pl}^{2/3}$. \footnote{Of
course, the parameter     $\bar{M}_{Pl} \sim G^{-1}$ is infrared-free.} At
the same time the holographic bound is alleviated in both models. 

Note
that the amount of running in the UV one can expect depends on the 
transfer of energy
between the various components at a particular cosmological epoch. In
particular, when $G$ is not static and owing to the Bianchi identity satisfied
by the Einstein tensor on the ${\sl lhs}$ of the Einstein equation, it is a 
quantity $GT_{total}^{\mu \nu}$,
where $T_{total}^{\mu \nu}$ is the total energy-momentum tensor, which is
actually conserved. Here $T_{total}^{\mu \nu}$ counts all forms of energy
densities (radiation, matter, vacuum, ...). Owing to the fact that the 
total energy-momentum tensor is no longer conserved when $G$ is running,
the energy transfer between various forms of energy densities is now
also controlled  by the scaling of the gravitational coupling. This  leads to the generalized equation of continuity, which 
in the FLRW metric takes the form 
\begin{equation}
\dot{G}(\rho_{\Lambda } + \rho_m ) + G \dot{\rho }_{\Lambda } +
G (\dot{\rho }_{m} + H\alpha_m \rho_m )  = 0  \;.
\end{equation}
Here overdots denote time derivative, $\alpha_m \equiv 3(1 + \omega_m)$
introduces the EOS for matter $p_m = \omega_m \rho_m$, and $p_{\Lambda}
\simeq -\rho_{\Lambda}$. 

At the stage of inflation it is naturally to set
$\rho_m \simeq 0$, so $G$ varies only at expense of the inflaton potential
energy. However, the slow-roll approximation precludes any  discernible 
running at this stage. Also, from this model-independent point of view,
it is impossible to fix the gradient of $G$.

This situation is however changed upon considering
the subsequent cosmological
epoch, where matter is being created. With canonical assumption that matter
is conserved, assuming also some leftover vacuum energy in the spirit of the
holographic bound (1), $\rho_{\Lambda}(\mu) \simeq c\mu^2 G^{-1}(\mu)$,
\footnote{This strongly resembles the model for holographic dark energy
\cite{28}, a stuff prevailing at present but suppressed at earlier
cosmological epochs.}
where $\mu$ is the energy scale and $c \lsim 1$, we obtain from (4) that
\begin{equation}
\mu = - \frac{G'(\mu ) \rho_m }{2c} \;.
\end{equation}
Some scaling properties of $G$ as implied by holography can be easily
inferred from (5). Namely, from the
requirement of the positivity of the scale $\mu $, $\mu > 0$, it is
seen that $G'(\mu ) < 0$, which consequently means that $\dot{G}(t)
> 0$, {\it i. e.}, $G(t)$ increases as a function of cosmic time. Such
a scale dependence implies that the coupling $G$ is asymptotically free;
a feature exhibited, for instance, by   higher-derivative quantum gravity
models at the 1-loop level \cite{29, 30, 31} or in conformal 
field theory in curved
spacetime \cite{32}. More generally, this fits with the approach 
to quantum gravity, based
on the existence of a nontrivial fixed point for the dimensionless
gravitational coupling in the UV (for a review see \cite{33}).
We find it very appealing to point to the
asymptotically free regime for $G$ in a quite model independent way, relying
solely on the holographic principle. Consequently, the holographic bound (1)
is alleviated in this regime.

In conclusion, we have noted that besides the restriction on the UV cutoff
coming from the scale of violation of the tree-level unitarity, the
effective field theory approach to the Higgs inflation should not also 
be used to describe states inside a black hole. Since the latter constraint
gives the upper limit on the UV cutoff, we have shown that both restrictions
are, in general, very hard to reconcile. We have noted that both 
restrictions can
be alleviated in the scenario of asymptotically safe gravity. We have
demonstrated, by relying on the holographic principle only and using a set of
reasonable assumptions, how this goal can be achieved in a model independent
way.

{\bf Acknowledgment. } This work was supported by the Ministry of Science,
Education and Sport
of the Republic of Croatia under contract No. 098-0982887-2872.


\begin{thebibliography}{160}
\bibitem{1} F. L. Bezrukov and M. Shaposhnikov, Phys. Lett. B659 (2008) 703.
\bibitem{2} F. Bezrukov, A. Magnin, M. Shaposhnikov and S. Sibiryakov,
arXiv:1008.5157 [hep-ph].
\bibitem{3} R. N. Lerner and J. McDonald, JCAP 1004 (2010) 015; Phys. Rev.
D82 (2010) 103525.
\bibitem{4} C. P. Burgees, H. M. Lee and M. Trott, JHEP 0909 (2009) 103.
\bibitem{5} J. L. F. Barbon and J. R. Espinosa, Phys. Rev. D79 (2009)
081302.
\bibitem{6} C. P. Burgees, H. M. Lee and M. Trott, JHEP 1007 (2010) 007.
\bibitem{7} M. Atkins and X. Calmet, arXiv:1002.0003 [hep-th].
\bibitem{8} M. Atkins and X. Calmet, arXiv:1005.1075 [hep-ph].
\bibitem{9} M. P. Hertzberg, arXiv:1002.2995 [hep-ph].
\bibitem{10} S. Ferrara, R. Kallosh and A. Linde, arXiv:1008.2942 [hep-th].
\bibitem{11} M. Buck, M. Fairbairn and  M. Sakellariadou,
Phys. Rev. D82 (2010) 043509.
\bibitem{12} J. L. Cervantes-Cota and H. Dehnen, Nucl. Phys. B442 (1995)
391; Phys. Rev. D51 395 (1995).
\bibitem{13} C. Germani and A. Kehagias, Phys. Rev. Lett. 105 (2010) 011302.
\bibitem{14} M. Atkins and X. Calmet, arXiv:1011.4179 [hep-ph].
\bibitem{15} G. F. Giudice and H. M. Lee, arXiv:1010.1417 [hep-ph].
\bibitem{16} F. Bauer and D. A. Demir, arXiv:1012.2900 [hep-ph].
\bibitem{17} A. Cohen, D. Kaplan and A. Nelson, Phys. Rev. Lett. 82 (1999)
4971.
\bibitem{18} J. Maldacena, Adv. Theor. Math. Phys. 2 (1998) 231.
\bibitem{19} G. 't Hooft gr-qc/9311026.
\bibitem{20} L. Susskind, J. Math. Phys. (NY) 36 (1995) 6377.
\bibitem{21} D. A. Lowe, J. Polchinski, L. Susskind, L. Thorlacius and J.
Uglum, Phys. Rev. D52 (1995) 6997.
\bibitem{22} See e.g.,  L. Susskind, J. Lindesay: ``An Introduction to
Black Holes, Information And The String Theory Revolution: The Holographic
Universe", World Scientific Publishing Company (2004). 
\bibitem{23} R. Horvat, Phys. Lett. B674 (2009) 1.
\bibitem{24} P. Frampton, S. D. H. Hsu, T. W. Kephart and D. Reeb,
Class. Quant. Grav. 26 (2009) 145005.
\bibitem{25} C. Amsler et al. [Particle Data Group], Phys. Lett. B667 (2008)
1.
\bibitem{26} F. Bezrukov, D. Gorbunov and M. Shaposhnikov, JCAP 0906 (2009)
029.
\bibitem{27} C. Germani and A. Kehagias, JCAP 1005 (2010) 019 [Erratum-ibid.
1006 (2010) E01]. 
\bibitem{28} S. D. Hsu, Phys. Lett. B594 (2004) 1; M. Li, Phys. Lett. B603
(2004) 1; R. Horvat, Phys. Rev. D70 (2004) 087301. 
\bibitem{29} J. Julve and M. Tonin, Nuovo Cimento B46 (1978) 137.
\bibitem{30} E. S. Fradkin and A. A. Tseytlin, Nucl. Phys. B201 (1982) 469.
\bibitem{31} A. O. Barvinsky, Phys. Lett. B159 (1985) 269.
\bibitem{32} J. Sola, J. Phys. A41 (2008) 164066.
\bibitem{33} R. Percacci, arXiv:0709.3851 [hep-th].

\end{thebibliography}
\end{document}